\documentclass[preprint,aps,pra,showpacs]{revtex4}
\usepackage{epsfig}

\begin{document}

\title{Analysis of Atom-Interferometer Clocks}

\author{Steven~Peil}
\email{steven.peil@usno.navy.mil}
\author{Christopher~R.~Ekstrom}
\affiliation{United States Naval Observatory, Washington, DC 20392}

\date{\today}

\begin{abstract}
We analyze the nature and performance of clocks formed by stabilizing an oscillator to the phase difference between two paths of an atom interferometer.  The phase evolution has been modeled as being driven by the proper-time difference between the two paths, although it has an ambiguous origin in the non-relativistic limit and it requires a full quantum-field-theory treatment in the general case. We present conditions for identifying deviations from the non-relativistic limit as a way of testing the proper-time driven phase evolution model. We show that the system performance belies the premise that an atom-interferometer clock is referenced to a divided-down Compton oscillation, and we suggest that this implies there is no physical oscillation at the Compton frequency.

\end{abstract}

\maketitle


\section{Introduction}

In a recent paper, Lan {\em et al.}~\cite{lan} claim to have created a Compton clock, an oscillator with frequency stabilized to a particle's mass via a purported intrinsic oscillation at the Compton frequency, $\omega_0=mc^2/\hbar$.  The ultimate limits of such a clock would far surpass any current or proposed traditional atomic clock; the experiment used $^{133}$Cs, for which $\omega_0/2\pi=3\times 10^{25}$~Hz, about 10 orders of magnitude higher than the frequency of optical atomic clocks, the current state-of-the-art.

Whether $\omega_0$ corresponds to a physical oscillation is a matter of debate.  De Broglie first associated a massive particle with an oscillation at frequency $\omega_0$ in developing his wave theory of matter, and it was integral to his pilot-wave theory of quantum mechanics~\cite{debroglie}. Despite the success of the theory of matter waves, the idea of an internal clock for a massive particle has been mostly ignored~\cite{bohm}.  But in recent years it has appeared in the literature with claims that the redshift of the Compton oscillation frequency is measured in an atomic gravimeter~\cite{compton_redshift}, as well as with a search for evidence of an internal oscillation in the electron in channeling experiments~\cite{tunneling}.  In a recent book, Penrose claims that a stable, massive particle behaves as a very precise quantum clock, ``oscillating'' at $\omega_0/(2\pi)$~\cite{penrose}.

Yet the overall phase of a quantum state, to which the Compton clock would fundamentally correspond, is widely accepted to be unobservable, while it is the relative phase between two states that corresponds to a physical oscillation. This is manifest in the squaring of the state vector characterizing the system to generate the probability distribution for measurement outcome, which at the same time does away with the overall phase.  In recent literature it has been claimed that $\omega_0$ does not correspond to a physical oscillation, with the point made that it is not Doppler shifted to an observer in motion~\cite{greenberger}, and that an atom interferometer is a single accelerometer, and not two clocks~\cite{SS}.

What seems to be accepted is that, in non-relativistic quantum mechanics, the Compton frequency is the rate of accumulation of the quantum phase with respect to proper time $\tilde{\tau}$ for a system. In an atom-interferometer clock, the difference of quantum phases of two motional states of an atom is used as a frequency reference~\cite{lan}. Here we analyze the source of phase evolution and frequency stability in this system, including the role of the Compton frequency and purported oscillation.

\section{Phase Evolution in an Atom Interferometer}

Over the past two decades atom interferometers have developed into high-precision tools for inertial measurements~\cite{gustav,snadden}, tests of the foundations of general relativity~\cite{equiv}, and determination of the ratio $h/m$ of Planck's constant to atomic mass~\cite{hm_cs,hm_rb}.  Laser-atom interactions used for the beam splitting process produce superpositions of atomic states with different velocities that travel along two physically displaced paths and are subsequently recombined. The resulting interference reveals the difference in phase accumulated along each path.

The free-propagation phase factor for an atom with velocity $v$ can be written
\begin{equation}
e^{i\phi}=e^{i\omega_0 \tilde{\tau}}=e^{i\omega_0 t/\gamma}, \label{e.factor}  
\end{equation}
where $\tilde{\tau}$ is the proper time for the atom and $\gamma=1/\sqrt{1-v^2/c^2}$ is the Lorentz factor.  This can be seen from the path integral formulation of quantum mechanics~\cite{lan}, or by substituting relativistic momentum and energy expressions into the phase factor for a massive-particle wave packet~\cite{greenberger}. The assumption in each case is a well-defined (classical) trajectory characterizing the particle's motion; a full quantum field-theoretic treatment is required when the system is sensitive to relativistic effects, as a result of sufficiently high velocity and/or precision. In this semiclassical limit, the free-propagation phase difference between the two paths in the interferometer evolves at a rate given by
\begin{equation}
\omega=\omega_0(\gamma^{-1}-1), \label{e.freq}
\end{equation}
for the case where the second path corresponds to zero velocity in the lab frame.  This can be treated as the frequency of a two-state system formed by the motional states in the interferometer.

When $v\ll c$, the Lorentz factor can be expanded as $\gamma^{-1}\approx 1 -(v^2/2c^2) - (v^4/8c^4) +\ldots$ To lowest order in $v/c$, the above phase factor reduces to
\begin{equation}
e^{i\phi}=e^{i\omega_0 t/\gamma}\approx e^{i\omega_0 t}e^{-i mv^2t/2\hbar }.
\end{equation}
In this case the Compton frequency comes in only through an overall phase common to each path, which cancels when phase differences are considered. For the remaining factor, the phase evolves at a rate determined by the atom's kinetic energy, and the rate at which the relative phase in the interferometer evolves is just $mv^2/(2\hbar)$. 
There is no longer a factor of $\omega_0$ in the phase evolution of the interferometer, and no relativistic proper time is involved.

Of course, this phase evolution rate can be written as $\omega_0 v^2/(2c^2)$, which has been called the time-dilated Compton frequency~\cite{lan}. But the same term can also be written as $\omega_{dB}/2$, where $\omega_{dB}$ is the (angular) de Broglie wave frequency, the frequency associated with the physical oscillation of a matter wave with momentum $mv$. In the low velocity limit, then, the phase evolution cannot unambiguously be attributed to proper-time difference, since the rate is indistinguishable from that driven by kinetic-energy. These two origins of phase evolution give differing rates for arbitrary velocities, where formally a full quantum field-theoretic formalism is required, with Eq.~(\ref{e.freq}) quantifying the rate for the proper-time driven model, and the expression for a kinetic-energy driven phase evolution differing from this by the sign of the exponent of $\gamma$.~\cite{history}

If an interferometer were sensitive enough to resolve the next term in the expansion of $\gamma^{-1}$, the form of the phase evolution rate could be empirically tested. This term, $(v^4/8c^4)$, gives a contribution to the rate of phase evolution of
\begin{equation}
\delta \omega=\omega_0 (v^4/8c^4),
\end{equation}
or as a fractional frequency,
\begin{equation}
\delta \omega/\omega = v^4/\left(8c^4(\gamma^{-1}-1)\right), \label{e.ave}
\end{equation}
which reduces to $v^2/4c^2$ to lowest order in $v/c$. Resolving this term in an interferometer would be evidence of quantum phase evolution due to relativistic proper time.

\section{Comparison to Atomic Clocks}

In order to assess the precision with which an atom interferometer can measure the evolution of the relative phase between the two paths, we develop an analogy to conventional atomic clocks.

Conventional passive atomic clocks operate by stabilizing the frequency of a local oscillator to an atomic resonance.  Atomic states with well-defined internal energies make up a two-state system that serves as the frequency reference for the local oscillator, and the achievable signal-to-noise ratio and quality factor $Q$ of the resonance dictate clock performance. The signal-to-noise is determined by the statistics of measuring the quantum states of the number of atoms in the system, $N_a$, and $Q=f/\Delta f$, the ratio of frequency to linewidth, which is often measurement-time limited.  In the standard quantum limit (SQL), the precision of the frequency reference improves with averaging time $\tau$ as
\begin{equation}
\sigma (\tau)^{-1} = Q \sqrt{N_a} \sqrt{\tau},
\end{equation}
where $\sigma$ is the Allan deviation used for characterizing clocks and oscillators~\cite{allan}. This gives the averaging time $\tau_f$ required for a clock to resolve a fractional frequency $\delta f/f$ as
\begin{equation}
\tau_f = \left( Q \sqrt{N_a} (\delta f/f) \right)^{-2}.
\end{equation}

\begin{figure}
\includegraphics[width=0.5\textwidth]{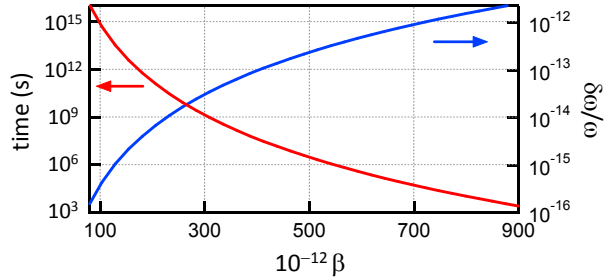}
\caption{(Color online.) Demonstration of conditions for an atom interferometer to be sensitive to relativistic effects.  For the parameters discussed in the text, the plot shows the fractional frequency, $\delta \omega /\omega$, representing the first order correction to the low velocity limit for phase evolution rate according to the proper time model (blue curve, right axis), and averaging time required to resolve this frequency (red curve, left axis), as a function of $\beta=v/c$.  For the system in Ref.~\cite{lan}, $\beta=100\times10^{-12}$.} \label{f.avetime}
\end{figure}

In an atom interferometer, the velocity states make up the two-state system serving as the frequency reference.  The quality factor is $Q = \omega_0 (\gamma^{-1}-1)/ \Delta \omega$, where the linewidth is determined by the interaction (Ramsey) time $T$, $\Delta \omega/(2\pi) \sim 1/2T$.  As an example, we consider a signal-to-noise ratio, $\sqrt{N_a}$, of 1000, corresponding to a sample size of order $10^6$ atoms, and a linewidth $\Delta \omega/(2 \pi) = 1.5$~Hz resulting from a 0.32~s measurement time.  For these values, we plot in Fig.~\ref{f.avetime} the time $\tau_{\omega}$ required to resolve the frequency difference $\delta \omega/\omega$, from Eq.~(\ref{e.ave}), as a function of $\beta = v/c$, assuming the SQL for all averaging times.  From the plot we see that at $\beta \sim 870 \times 10^{-12}$ ($v=0.26$~m/s), phase evolution due to relativistic proper time can be resolved with about 1 hour of averaging.  For cesium, this velocity corresponds to about 70 photon recoils~\cite{recoil}, which can be reached with state-of-the-art beam splitting techniques, so testing phase evolution due to relativistic proper time beyond the non-relativistic limit should be possible with current technology~\cite{kasevich100}.

The expected SQL for the system in Ref.~\cite{lan} can be determined from the experimental parameters.  The velocity difference between atoms in the two paths of the interferometer is due to the recoil from 10 photons, giving $v \approx .035$~m/s and $\omega/(2\pi) \approx 200$~kHz. The 320~ms measurement time and $10^6$ atoms contributing to the interferometer signal correspond to $\Delta \omega /(2 \pi) \approx 1.5$~Hz and a signal-to-noise of 1000.  So the expected performance for this system is characterized by an Allan deviation versus averaging time of $\sigma(\tau) = 2.5 \times 10^{-9}/\sqrt{\tau}$.  This is consistent with the standard-quantum-limited performance referred to in~\cite{lan}, and it corresponds to an averaging time of more than $7 \times 10^{14}$~s, or $2\times 10^7$ years, to test proper-time driven phase evolution, as can be seen from Fig.~\ref{f.avetime}. The performance actually demonstrated was two orders of magnitude worse than the SQL, making the averaging time required $7 \times 10^{18}$~s, or $2 \times 10^{11}$ years (more than 10 times the age of the universe). So that interferometer was not sensitive to relativistic effects, and the phase evolution was indistinguishable from that driven by the difference in kinetic energy due to absorption and emission of photons in the beam-splitting process.

We can also assess the potential performance of interferometer clocks in comparison to conventional atomic clocks. In most state-of-the-art atomic clocks the linewidth is measurement-time limited, so comparing $Q$s comes down to comparing frequencies. For microwave (fountain) clocks, the number of atoms will be comparable to an interferometer system, so the difference in quantum-limited performance is due only to the two frequencies.  For an interferometer clock to achieve stability comparable to a microwave clock with angular frequency $\omega_{\mu}$, it is required that $\omega_0(\gamma^{-1}-1) \sim \omega_{\mu}$.  For cesium, $\omega_0/(2\pi) = 3 \times 10^{25}$~Hz and $\omega_{\mu}/(2\pi) = 9.2 \times 10^9$~Hz; it is necessary for $(\gamma^{-1}-1) \approx v^2/(2c^2)$ to be on order of $3\times 10^{-16}$, giving $v/c \sim 2.5 \times 10^{-8}$ and $v \sim 7.4$~m/s. Beam-splitting for atom interferometers can produce relative velocities of roughly $100$ photon-recoil velocities, or about 0.3~m/s for cesium, and it may be possible to extend this another order of magnitude~\cite{kasevich100}.  This potential improvement could make the quantum-limited performance similar to microwave clocks.  In order to achieve atom-interferometer performance comparable to optical clocks, another several orders of magnitude in frequency would be required. Continuing to use the nonrelativistic approximation, it would be necessary to generate velocities $v \sim 2 \times 10^3$~m/s to achieve a frequency of $5\times 10^{14}$~Hz.  This is well beyond current technology; in addition, for a reasonably sized interaction region, such high velocities limit the measurement time and reduce the effective $Q$.

\section{Atom Interferometer Clock and Compton Oscillation}

In Ref.~\cite{lan}, the velocity states in an atom interferometer were used as a two-state frequency reference for a clock.  If the resulting clock were referenced to the purported Compton oscillation, it would suggest the promise of a reference with a frequency of $10^{25}$~Hz and correspondingly high $Q$.

The total phase determining the interferometer signal is the combination of the free propagation phase, discussed above, and the ``laser-atom'' phase from the beam-splitting interactions.  The interferometer in Ref.~\cite{lan} was turned into a clock by feeding back to the laser frequencies used in the experiment to adjust the laser-atom interaction and keep the total phase zero. Adjusting the laser frequencies would normally modify the reference frequency via the effect on the atom's velocity, which is entirely due to photon recoil.  An arrangement was used  to decouple the atom's velocity from the raw laser frequency $\omega_L$ by imposing $\omega_L=N \omega_m$, where $\omega_m=\omega_+ - \omega_-$ is the difference in frequencies of the beams used to impart velocity to the atoms, and $N$ is not necessarily an integer.  The atomic velocity then depends not on $\omega_L$ but on $N$, which is determined by a set of frequency multipliers. The physical output of the clock comes from adjusting an oscillator that determines $\omega_m$ (and $\omega_L$) to keep the total phase of the interferometer constant (=0). According to the paper, when $\phi_{\rm total}=0$, $\omega_m=\omega_0/(2nN^2)$.

In the previous section we analyzed the performance of an atom-interferometer clock based on the SQL for the two-state system comprised of different motional states separated in phase evolution rate by $\omega$.  The SQL for a true Compton clock would be governed by the frequency of the purported Compton oscillation, $10^{20}$ times higher. This same performance applies even to a system referenced to a divided down Compton oscillation, as claimed for the interferometer in Ref.~\cite{lan}.  This dividing down is reminiscent of optical clocks, where the optical frequency is divided down to an electronically countable microwave frequency.  The performance of the system is still determined by the optical frequency; the high stability associated with the optical signal is transferred without degradation to the lower-frequency oscillation. This is simply a consequence of propagation of uncertainties, which shows that ``dividing down'' the frequency of an oscillator does not change the relative uncertainty, {\em i.e.} stability, of the signal~\cite{unc_prop}.

If the frequency of the interferometer clock were divided down from the frequency of a physical oscillation at $\omega_0$ in the same way, the stability of the clock should be the same as that oscillation~\cite{dw},
\begin{equation}
\delta \omega / \omega = \delta \omega_0 /\omega_0.
\end{equation}
The observation that $\omega$ is derived from the difference of two frequencies, the phase evolution rates of the arms of the interferometer, each of which could be considered a divided-down Compton frequency, does not change this. Because the two rates are both fractions of the same higher frequency, the correlations introduce a non-zero covariance, with uncertainty propagation again showing the final stability to be the same as that of the oscillation at $\omega_0$~\cite{unc_prop2}. On the other hand, if the covariance between the two frequencies is zero, indicating no correlation, the derived stability agrees with that calculated in the previous section.

The only qualification to this argument is that the factor in $\omega$ multiplying (dividing down) the Compton frequency, $\eta = \gamma^{-1}-1$, has an associated uncertainty that needs to be included in the error propagation and that would limit the clock performance in this model. This technical-noise limit should be independent of any standard quantum limit, so the predictions of clock performance from a Compton-oscillation model can still be differentiated from a clock governed by the properties of the two-velocity-state system, and the empirical stability provides a means of testing the physical reality of the Compton oscillation. We can try to assess the technical-noise limit in Ref.~\cite{lan} from experimental values provided.  In the low-velocity limit, uncertainty propagation from the Compton oscillation to the clock output includes the term $\Delta \eta / \eta = 2\Delta v/v=4 \Delta N/N$.  An uncertainty on $N$ is not stated, but from the number of significant digits it can be inferred that $\Delta N/N \sim 10^{-12}$~\cite{prec_N}.  The Compton oscillation model therefore predicts a technical-noise limited stability no worse than $10^{-12}$, significantly better than that predicted by the two-velocity-state SQL and better than that demonstrated.

So the system in~\cite{lan}, despite being designed to operate as a divided down Compton clock, empirically operates as a clock with no connection to the $Q$ of the purported Compton oscillation.  The ``Compton oscillations'' for an atom in the two paths of the interferometer behave as completely uncorrelated, even though affiliated with the same particle.  This evidence strongly suggests that the frequency $\omega_0$ does not correspond to a physical oscillation.  In the absence of a physical oscillation, what remains is the quantum phase, which is just a product of a the system's proper time and the parameter $\omega_0$. The phase still accumulates at a high rate (compared to frequencies that we are accustomed to for atomic clocks) due to the large value of $\omega_0$, but this does not contribute to making a clock until a relative phase is generated. In that case, for common velocities, the rest-mass contribution mostly cancels, and the resulting frequency is not exceptional~\cite{v_c}.

Finally, we point out that, if these conclusions stand, there may be an impact on past analysis of an atom interferometric measurement of the gravitational redshift and a corresponding test of Local Position Invariance (LPI).  In Ref.~\cite{compton_redshift}, a measurement of $g$ was re-interpreted as a measurement of the gravitational redshift of the Compton frequency and a test of the dependence of the redshift on gravitational potential $U$.  If the notion of a physical Compton oscillation is abandoned, the implications for these results is unclear.  The nature of the measurement changes, from a measurement of the redshift (of the frequency of a physical oscillation) to a measurement of time dilation (due to the geometry of spacetime).  In particular, it is not obvious that the anomalous redshift parameter $\beta$ used in~\cite{compton_redshift} would enter in the same way. This may be a topic for future consideration.

\section{Conclusions}

We have analyzed the nature and performance of clocks based on atom interferometers. We have presented the conditions required to test proper-time driven phase evolution for general conditions beyond the non-relativistic limit.  We have shown that atom-interferometer clocks are not competitive with state-of-the-art atomic clocks; in particular, performance comparable to optical clocks seems to be out of reach. This modest performance is despite claims that the system is a Compton clock, referenced to a purported oscillation at $\omega_0$, which suggests possible performance 10 orders of magnitude better than optical clocks. This discrepancy, illustrated using propagation of uncertainties from the presumed Compton oscillation to the clock output, strongly suggests that there is no physical oscillation at $\omega_0$.

\section{Acknowledgements}

We acknowledge useful discussions with J.~Hanssen.

\bibliographystyle{prsty}

\begin{thebibliography}{10}
\bibitem{history}
Considering the difference in these two phase evolution rates was instrumental in the original development of de Broglie's matter-wave theory.
\bibitem{recoil}
A ``photon recoil'' is the recoil velocity of an atom from absorbing or emitting a single photon.
\bibitem{unc_prop}
If $\omega_{\mu}$ is the lower  microwave frequency signal, $\omega_{\rm opt}$ is the optical frequency that is stabilized, and $M$ is a large integer (usually on order of $10^5$) relating the two, $\omega_{\mu}=\omega_{\rm opt}/M$, then $\delta \omega_{\mu}=\delta \omega_{\rm opt}/M$, and $\delta \omega_{\mu}/\omega_{\mu}=\delta \omega_{\rm opt}/(M \omega_{\mu})=\delta \omega_{\rm opt}/\omega_{\rm opt}$.
\bibitem{unc_prop2}
From Eq.~(\ref{e.freq}) we have $\omega = \gamma^{-1} \omega_0 - \omega_0$.  Uncertainty propagation for the absolute uncertainties when subtracting two values gives
\begin{equation}
\left( \delta \omega \right)^2 = \left(\gamma^{-1} \delta \omega_0 \right)^2 + \left(\delta \omega_0  \right)^2 - 2 \gamma^{-1} \delta_{0,0} ,
\end{equation}
where $\delta_{0,0}$ is the covariance of $\omega_0$ with itself, {\em i.e.} $(\delta \omega_0)^2 $.  The expression for the absolute uncertainty reduces to $\delta \omega = \delta \omega_0 (\gamma^{-1}-1)$, and the relative uncertainty is then $\delta \omega / \omega = \delta \omega_0 (\gamma^{-1}-1)/(\omega_0(\gamma^{-1}-1)) = \delta \omega_0 /\omega_0$.  Propagation of uncertainties shows that not only does dividing down the frequency of an oscillator not affect the stability, but differencing two frequencies divided down from the same oscillator does not affect the stability; in each case, the final signal should exhibit the same stability as the high-frequency oscillation.
\bibitem{v_c}
At first glance, it seems that the system could exhibit performance consistent with the notion of a Compton clock for large enough velocities to make $\omega \rightarrow \omega_0$, which requires $v \rightarrow c$. But, in addition to being an unrealistic limit, the non-relativistic formalism used to derive the phase evolution in the interferometer is not valid for relativistic velocities and must be replaced with a full quantum field-theoretic treatment in that case~\cite{feynman_hibbs}, as we have emphasized all along.
\end{thebibliography}

\end{document}